\documentclass[reprint,prd,onecolumn,notitlepage,11pt,nofootinbib]{revtex4-1}
\usepackage[english]{babel}
\usepackage{amsmath,amssymb,graphicx,bm,lipsum,float}
\usepackage[colorlinks=true,citecolor=blue,linkcolor=blue,urlcolor=blue]{hyperref} 
\usepackage[font={small},flushleft,indent]{caption}
\usepackage{setspace}

\usepackage{subcaption} 
\usepackage{placeins}
\usepackage{overpic}

\begin{document} 
\title{The Terrell-Penrose Effect Revisited and Extended for Black Hole Spacetime}

\author{Qing-Hua Zhu}
\email{zhuqh@cqu.edu.cn}
\affiliation{Department of Physics and Chongqing Key Laboratory for Strongly Coupled Physics, Chongqing University, Chongqing 401331, China}

\begin{abstract}
Half a century ago, Terrell and Penrose independently studied the visual appearance of a moving sphere and found that the visual distortion can be formulated by a conformal transformation on the observer's celestial sphere. 
In this study, we examine the Terrell-Penrose effect by considering the images of spheres in two contrasting cases: i) a static sphere in the view of a moving observer, and ii) a moving sphere in the view of a static observer. In these two cases, the images are expected to be identical for flat spacetime because of the principle of relativity. However, utilizing ray-tracing simulations, our results find that the images are distinct, and in case ii) the image distortion violates conformality. 
We also extend the studies to Schwarzschild spacetime. In case i), we can obtain a natural extension, whereas in case ii), it is demonstrated that gravity introduces an additional influence on the images of moving objects.   
\end{abstract}

\maketitle 

\section{Introduction}

The first images of the supermassive black holes at the centers of M87 and Sgr A* \cite{EventHorizonTelescope:2019dse,EventHorizonTelescope:2022wkp} 
have renewed the theoretical interest in horizon-scale emission sources, such as accretion disks and relativistic jets \cite{EventHorizonTelescope:2021srq,Luminet:1979nyg,Jaroszynski:1997bw,Bisnovatyi-Kogan:2022ujt,Igata:2025glk,deSa:2024dhj}. 
Since light rays emitted from a source near a black hole propagate along geodesics to a distant observer as described by Einstein's gravity, the resulting image is determined collectively by the emission configuration, gravitational light deflection, and relativistic aberration. The emission configuration arises from the dynamics of the accreting matter, whereas the latter two stem from relativistic  effects.   
Imaging a black hole influenced by the aberration effect is of theoretical interests \cite{Grenzebach:2015uva,Stuchlik:2018qyz,Briozzo:2022mgg,Zhu:2023kei}, and used in the fiction movie \cite{James:2015yla}.  
In astrophysical observations, this aberration effect on the images can not be neglected when taking into account an observer comoving with the expansion of the universe \cite{Lebedev:2013qoa,Lebedev:2016kun,Perlick:2018iye,Qi:2019zdk,Chang:2019vni,Tsupko:2019mfo,Li:2020drn,Chang:2020miq,He:2020dfo,Guo:2022nto}.   
As the angular resolution of future very-long-baseline interferometry (VLBI), such as the ngEHT \cite{Zhang:2024owe} and BHEX \cite{Johnson:2024ttr}, continues to improve, imaging more distant supermassive black holes will become feasible. In the far future, this might provide an opportunity to detect the aberration effect induced by comoving observer situated in an expanding universe.

\begin{figure} \centering
  \includegraphics[width=0.45\linewidth]{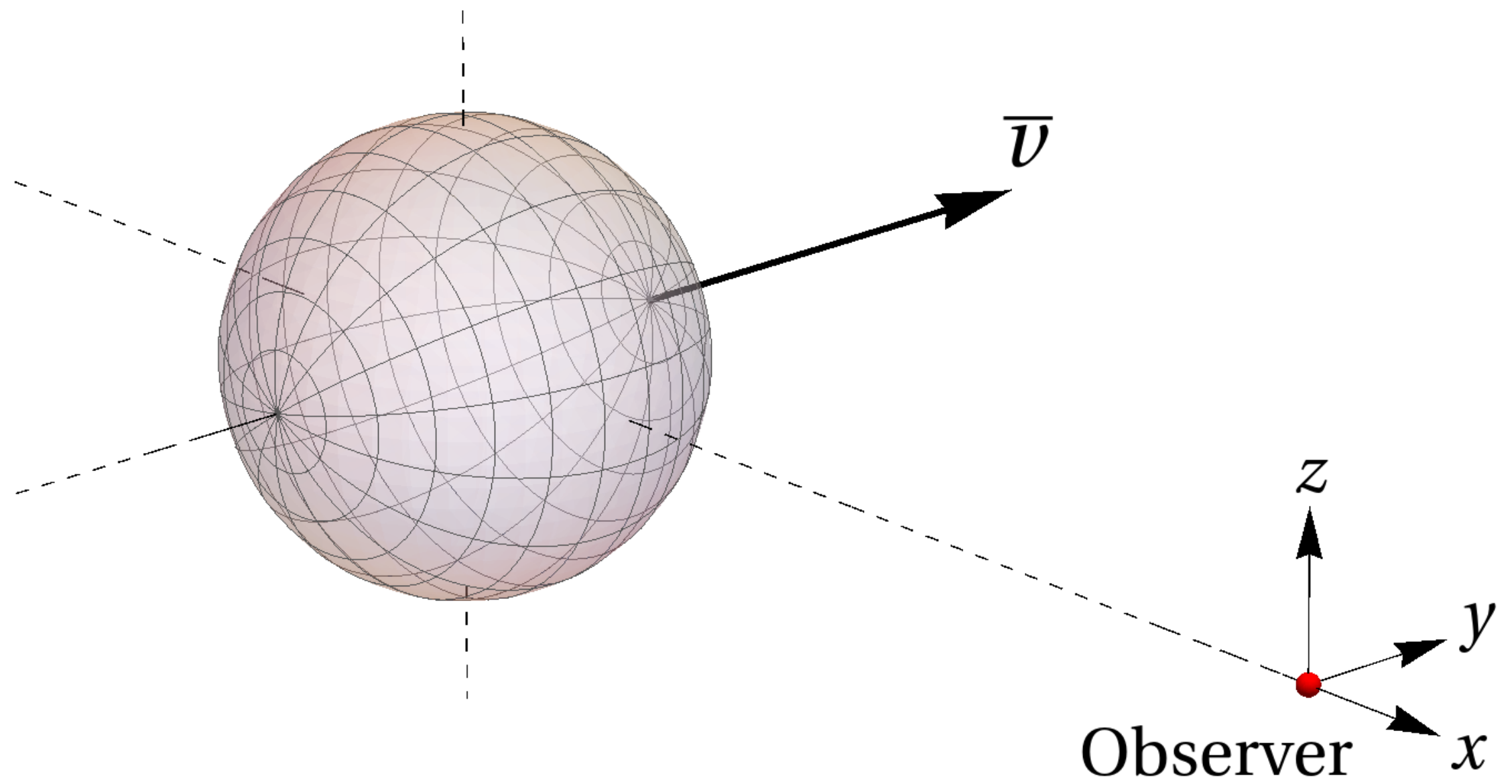}
  \caption{The moving sphere seen by the observer. The observer's line of sight (to the sphere's center) and the sphere's direction of motion are perpendicular.\label{F0}}
\end{figure}

In the late 1950s, Terrell and Penrose presented an insightful result on the aberration effect in special relativity, so-called Terrell-Penrose effect \cite{Terrell:1959zz,Penrose:1959vz,PhysRev.161.1295,10.1119/1.16131}. They independently found that the Lorentz contraction of a moving sphere is invisible to the observer. Instead, optical appearance of the sphere remains circular as its speed increases. 
 The schematic diagram for the moving sphere is shown in Fig.~\ref{F0}. 
In the seminal papers \cite{Terrell:1959zz,Penrose:1959vz}, the main result is the conformal transformation describing the image distortion on the celestial sphere,  which was derived from the relativistic aberration formula \cite{Einstein:1905ve},
\begin{eqnarray}
  \cos\theta' = \frac{\cos\theta - \bar{v}}{1-\bar{v}\cos \theta}~. \label{1}
\end{eqnarray}
Eq.~(\ref{1}) arises from the Lorentz transformation and relates two angular distances $\theta'$ and $\theta$ defined in two inertial frames with relative speed $\bar{v}$. Based on the principle of relativity, it was claimed that Eq.~(\ref{1}) could also be used to formulate the visual distortion of a moving sphere seen by a static observer \cite{Terrell:1959zz,Penrose:1959vz}. 
On the other hand, both physical analysis and numerical simulations showed that the image of a moving object exhibits an apparent rotation \cite{Terrell:1959zz,PhysRev.161.1295,10.1119/1.16131,Hornof:2024rzt}.  Recently, the predicted optical appearance of a moving sphere has been tested experimentally, where apparent rotation was detected using the laser pulses and ultra-fast photography  \cite{Hornof:2024rzt}. 
However, if the sphere seen by the observer shown in Fig.~\ref{F0} is rotated, it can no longer be explained by the conformal transformation formulated by Terrell and Penrose \cite{Terrell:1959zz,Penrose:1959vz}. There is an apparent discrepancy that requires clarification. 

In this study, we revisit the Terrell-Penrose effect by explicitly distinguishing between two contrasting cases: i) a static object in the view of a moving observer, and ii) a moving object in the view of a static observer, using the relativistic ray-tracing simulations developed in Ref.~\cite{Zhu:2024vxw}.  In case i), the observer's celestial sphere is established adopting an operationalist approach, for the sake of naturalness, following Refs.~\cite{Chang:2020miq,Zhu:2023kei}. 
It thus allows a direct extension to curved spacetime. In case ii), we use time-series reconstruction to perform slow-light ray tracing for moving emission sources.
The `slow-light' approach includes the propagation time of light in ray tracing, in contrast to the `fast-light' approximation \cite{White:2022paq,Cardenas-Avendano:2024sgy}.  It ensures that the image is constructed from light rays received simultaneously by the observer.  In flat spacetime, the apparent distortion of the images originates from two distinct effects: Lorentz contraction and apparent elongation from the slow-light effect. 
Additionally, we also present a tentative extension of these situations to black hole spacetime. 
In the presence of gravity, the images produced in cases i) and ii) could not be identical, as the emission sources and the observers could not be situated in the same gravitational environment. Thus, it could be a crucial way to decode spacetime geometry.  
Here, we are particularly interested in the apparent elongation additionally influenced by gravity. 

For illustration, we introduce the length $l_g$ that characterizes the scale of the emission sources and the observer's distance to the black hole. In the regime where $l_g$ is comparable to the gravitational radius $r_g$, gravitational effects become significant and can not be ignored. In contrast, when $l_g\gg r_g$, the influence of gravity is negligible. Throughout this paper, we regard the regime $l_g= 10^{4} r_g$ as the flat spacetime limit.   
 
The rest of the paper is organized as follows. In Sec.~\ref{II}, we briefly review the ray-tracing simulations and the construction of the celestial sphere using reference light rays. In Sec.~\ref{III}, we derive the conformal transformation induced by the aberration effect in Schwarzschild spacetime, and show the apparent distortion of spheres and disks. In Sec.~\ref{IV}, we present the time-series reconstruction scenario for imaging moving extended bodies, and study the apparent distortion in the absence and presence of gravity. In Sec.~\ref{V}, we present a comparison between the cases of a static sphere seen by a moving observer and a moving sphere seen by a static observer in flat spacetime. In Sec.~\ref{VI}, we summarize the conclusions and discussions.

\section{Brief review of ray-tracing simulations and celestial coordinates \label{II}}

The key feature of ray-tracing simulations in black hole spacetime is bending of light by gravity. A spherical spacetime can be formulated by the metric as follows, 
\begin{equation}
  \textrm{d}s^2 = -\frac{\textrm{d}t^2}{f(r)}+f(r)\textrm{d}r^2+r^2(\textrm{d}\theta^2+\sin^2\theta\textrm{d}\phi^2)~,
\end{equation}
where $f(r)=1-2M/r$ for Schwarzschild black hole and the gravitational radius is $r_g\equiv M$. Based on the metric, the 4-momentum of light rays is given by
\begin{eqnarray}
  k&=&E\Bigg( -\textrm{d}t\pm_r \frac{1}{f}\sqrt{1-\frac{\rho^2 f}{r^2}}\textrm{d}r  \pm_\theta \rho \sqrt{1-\cos^2\varphi \left( \frac{\sin\theta_\text{o}}{\sin\theta}\right)^2}\textrm{d}\theta  - \rho\cos\varphi\sin\theta_\text{o} \textrm{d}\phi\Bigg)~,   \label{mom}
\end{eqnarray}
where $\pm_r$ and $\pm_\theta$ are the plus-minus signs, $\theta_\text{o}$ is inclination for observers and $E$ is an integral constant representing the energy of light. Here, we have rewritten the other integral constants (angular momentum and Carter constant) in terms of $\theta_\text{o}$, $\rho$ and $\varphi$, namely, $\rho \equiv \sqrt{K}/E$ and $\cos\varphi\equiv-L/(\sqrt{K}\sin\theta_\text{o})$  \cite{Zhu:2024vxw,Zhu:2025jqh}. In black hole spacetime, the light rays emitted from a source can reach the observer multiple times, resulting in primary, secondary, and $n$-th order images \cite{Luminet:1979nyg,Virbhadra:1999nm,Virbhadra:2008ws}. For each image order, it showed that only one light ray connects the location of a source $\bm x_\text{s}[=(r_\text{s}, \theta_\text{s}, \phi_\text{s})]$ and the location of an observer $\bm x_\text{o}[=(r_\text{o}, \theta_\text{o}, \phi_\text{o})]$ \cite{Zhu:2024vxw}, except when the source is situated on the caustic point. Specifically, we have one-to-one mapping between the $(\bm x_\text{s}, \bm x_\text{s})$ and the impact parameters of light rays, namely,
\begin{eqnarray}
  \text{RT}:(\bm x_\text{o}, \bm x_\text{s},n) 	\mapsto (\varphi,\rho)~. \label{RT}
\end{eqnarray}
The ray-tracing simulations used in this study is completely described by Eq.~(\ref{RT}).

\begin{figure} \centering   
  \centering
  \includegraphics[width=0.5\linewidth]{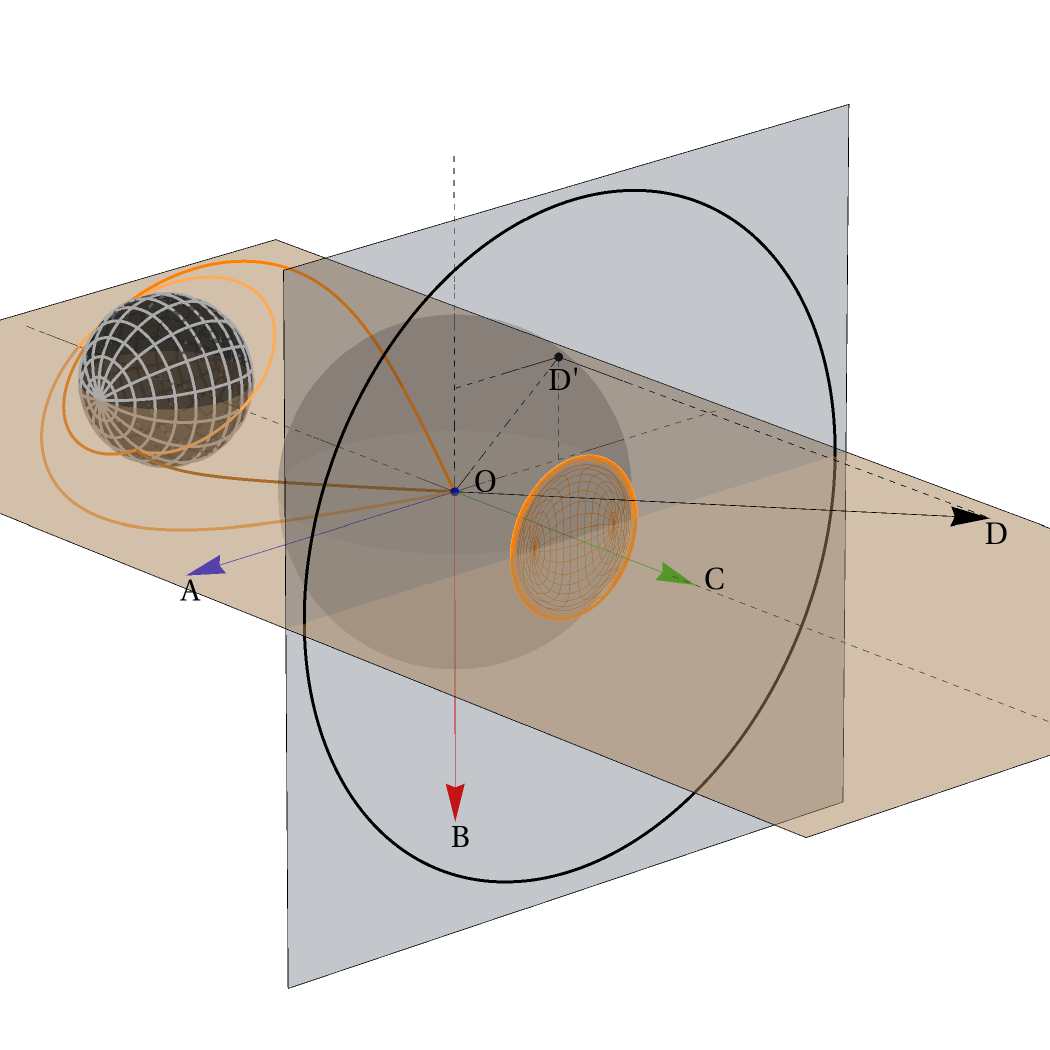}
  \caption{Schematic diagram of imaging a sphere onto projection plane.  The $\overrightarrow{\rm OA}$, $\overrightarrow{\rm OB}$ and $\overrightarrow{\rm OC}$ represent the spatial components of the vectors $k^{(\phi)}$, $k^{(\theta)}$ and $k^{(r)}$, respectively. We define a celestial sphere centered at point O, using a coordinate system where the $x$-, $y$-, and $z$-axes are OA, OB, and OC, respectively.  Light ray $\widetilde{\rm SO}$ propagates from a point source S to the observer O. The direction vector of the light ray $\widetilde{\rm SO}$ at location O is $\overrightarrow{\rm OD}$. The $\overrightarrow{\text{OD}'}$ represents the projection on plane AOB. The celestial coordinates for locating the light ray $\widetilde{\rm SO}$ are defined as $\Phi_0\equiv \angle\text{AOD}'$ and $\Psi_0\equiv\angle\text{COD}$.  We use Lambert azimuthal projection to map celestial coordinates $(\Phi_0,\Psi_0)$ onto image plane denoted as $(\text{X},\text{Y})$. 
  \label{F1}} 
\end{figure}

Following an operationalist perspective, the celestial coordinates of a light ray received by an observer are established relative to a set of reference light rays \cite{Chang:2020miq,Chang:2020lmg,Chang:2021ngy,He:2020dfo,Guo:2022nto}, which can be given by \cite{Zhu:2024vxw},
\begin{subequations}
  \begin{eqnarray}
    k^{(r)}&=&k\big|_{\rho=0,\pm_r=+}~,\\
    k^{(\theta)}&=&k\big|_{\rho=r_\text{o}/\sqrt{f(r_\text{o})},\varphi=\pi/2,\pm_r=\pm_\theta=+}~,\\
    k^{(\phi)}&=&k\big|_{\rho=r_\text{o}/\sqrt{f(r_\text{o})},\varphi=0,\pm_r=\pm_\theta=+}~,
  \end{eqnarray} \label{refk}
\end{subequations}
where the 4-momentum $k$ has been given in Eq.~(\ref{mom}).
The schematic diagram for the ray tracing and the celestial sphere is shown in Fig.~\ref{F1}.
The angular distance between 4-vectors $k_1^\mu$ and $k_2^\mu$ on observer's sky can be derived from the inner product as follows,
  \begin{eqnarray}
     \langle k_1, k_2 \rangle & \equiv & \frac{\gamma^{\ast} k_1}{|
    \gamma^{\ast} k_1 |} \cdot \frac{\gamma^{\ast} k_2}{|
    \gamma^{\ast} k_2 |}  = \left(1+\frac{k_1\cdot k_1}{(u\cdot k_1)^2}\right)^{-\frac{1}{2}} \nonumber \\ && \times \left(1+\frac{k_2\cdot k_2}{(u\cdot k_2)^2} \right)^{-\frac{1}{2}}\left(1 + \frac{k_1 \cdot k_2}{(u \cdot k_1) (u \cdot
    k_2)}\right)~, \label{defAngle} \nonumber \\ 
  \end{eqnarray} 
where the angular distance is $\angle(k_1,k_2)\equiv  \arccos\langle k_1,k_2\rangle$, the `$\cdot$' represents the metric contraction,  $|\gamma^\ast k|\equiv \sqrt{\gamma^\ast k\cdot \gamma^\ast k}$, $\gamma^\ast k^\mu \equiv \gamma^\mu_\nu k^\nu$, and $\gamma^\mu_\nu[=\delta^\mu_\nu +u^\mu u_\nu]$ is spatial projection operator. Here, the $k_1$ and $k_2$ are not necessary null vectors.
Using Eqs.~(\ref{refk}) and Eq.~(\ref{defAngle}), one can verify that at the position $\bm x_\text{o}$ of a static observer with 4-velocity $u^\text{(stc)}  =  f(r_\text{o})^{-1/2} \partial_t$, the 4-momenta $k^{(r) }$, $k^{(\theta) }$ and $k^{(\phi) }$  are mutually orthogonal. Namely, we have $\langle k^{(\phi)},
k^{(r)} \rangle_\text{stc}$= $\langle k^{(\phi)}, k^{(\theta)} \rangle_\text{stc}$= $\langle
k^{(\theta)}, k^{(r)} \rangle_\text{stc} = 0$, where $\langle ...\rangle_\text{stc}\equiv \langle ... \rangle|_{u=u^\text{(stc)}}$. Thus, the celestial coordinates  $(\Phi_0, \Psi_0)$ can be introduced via
\begin{subequations}
  \begin{eqnarray}
  \cos \Phi_0 \sin \Psi_0&=& \langle k,k^{(\phi)}\rangle_\text{stc}~, \\
\sin \Phi_0 \sin \Psi_0 &=& \langle k,k^{(\theta)}\rangle_\text{stc}~,\\
\cos\Psi_0 &=& \langle k,k^{(r)}\rangle_\text{stc}~.
\end{eqnarray}  \label{CS}
\end{subequations}
Explicitly, we have
  \begin{eqnarray}
    \Phi_0  =   \varphi ~, & &
    \Psi_0  =   \arccos  
    \left(\pm\sqrt{1 - \frac{\rho^2 f_\text{o}}{r_\text{o}^2}} \right)  ~. \label{Image}
  \end{eqnarray} 
where we have denoted $f_\text{o}\equiv f(r_\text{o})$. A light ray, specified by  $(\rho, \varphi)$, can determine celestial coordinate $(\Phi_0, \Psi_0)$.

\section{Relativistic aberration effect \label{III}}

The celestial coordinates in Eq.~(\ref{Image}) are established for a static observer. To study the relativistic aberration, we will extend the formalism in Sec.~\ref{II} to the cases for observers in motion. In this study, we derive the aberration effect from an operationalist perspective, ensuring a systematic foundation rather than relying on ad hoc assumptions. 

\subsection{Aberration formulas \label{IIIA}}

Due to the aberration effect, the reference light rays $k^{(r)}$, $k^{(\theta)}$ and $k^{(\phi)}$ are no longer mutually orthogonal in the view of observers in motion. Specifically, product $\langle...\rangle$ of null vectors $k_1$ and $k_2$ can be generally evaluated to be  
\begin{eqnarray}
  \langle k_1, k_2 \rangle & = & {1 + \frac{\langle k_1, k_2
  \rangle_{\text{stc}} - 1}{(1 + z_1) (1 + z_2)}}~, \label{6}
\end{eqnarray}
where the Doppler redshift factor is given by $1 + z_\ast \equiv {(u \cdot k_\ast)}/{(u^\text{(stc)} \cdot k_\ast)}$ for $\ast=1$, $2$ and the $u$ is the 4-velocity of a moving observer. 
For example, substituting the $k_1$ and $k_2$ with $k^{(r)}$ and $k^{(\phi)}$, respectively, we obtain $\langle k^{(r)}, k^{(\phi)} \rangle = 1-1/((1+z^{(r)})(1+z^{(\phi)}))\neq0$. Namely, the $k^{(r)}$ and $k^{(\phi)}$ are no longer orthogonal for moving observers. 
Therefore, a set of new reference light rays is necessary and can be given by
\begin{subequations}
  \begin{eqnarray}
  k^{(r')} & = & k^{(r)} ~,\\
  k^{(\phi')} & = & k^{(\phi)} - \frac{\gamma^{\ast} k^{(\phi)} \cdot
  \gamma^{\ast} k^{(r')}}{| \gamma^{\ast} k^{(r')} |^2} k^{(r')}~,
  \\
  k^{(\theta')} & = & k^{(\theta)} - \frac{\gamma^{\ast} k^{(\theta)}
  \cdot \gamma^{\ast} k^{(r')}}{| \gamma^{\ast} k^{(r')} |^2}
  k^{(r')} \nonumber \\ && - \frac{\gamma^{\ast} k^{(\theta)} \cdot \gamma^{\ast}
  k^{(\phi')}}{| \gamma^{\ast} k^{(\phi')} |^2} k^{(\phi')} ~.
\end{eqnarray} \label{lightk}
\end{subequations}
By making use of these reference light rays, we define celestial coordinates $(\Phi,\Psi)$ for moving observer via 
\begin{subequations}
  \begin{eqnarray}
  \cos \Phi \sin \Psi  \equiv   \langle k, k^{(\phi')} \rangle~, \\
  \sin \Phi \sin \Psi  \equiv   \langle k, k^{(\theta')} \rangle~,\\
  \cos \Psi  \equiv   \langle k, k^{(r')} \rangle~.
\end{eqnarray}\label{11}
\end{subequations}
Detail expressions for Eqs.~(\ref{lightk}) and Eqs.~(\ref{11}) are presented in Appendix~\ref{AppA}.

Specifically, we consider a rotating observer with 4-velocity, 
  $u^{(\phi)}  =  {(\partial_t  + \omega
  \partial_{\phi})}/{\sqrt{f_\text{o}  - \omega^2 r_\text{o}^2 \sin^2
  \theta_\text{o}}}$,
where the $\omega[\equiv u^\phi/u^t]$ is angular speed. Associating the 4-velocity with Eqs.~(\ref{Image})--(\ref{lightk}), we derive the aberration formula relating celestial coordinates $(\Phi,\Psi)$ and $(\Phi_0,\Psi_0)$, namely,
\begin{subequations}
  \begin{eqnarray}
  \Phi & = & \arctan \left( \frac{f_\text{o} \sin \Phi_0 \sin \Psi_0}{\sqrt{f_\text{o}}
  \omega  r_\text{o}\sin \theta_\text{o} (1 - \cos \Psi_0) + f_\text{o} \cos \Phi_0 \sin \Psi_0}
  \right)~,  \label{9a} \\
  \Psi & = & \arccos \left( 1-\frac{(1 - \cos \Psi_0) (f_\text{o}   - \omega^2  r_\text{o}^2\sin^2  \theta_\text{o})}{ f_\text{o}+ \sqrt{f_\text{o}} \omega  r_\text{o}\sin  \theta_\text{o} \cos \Phi_0 \sin \Psi_0 } \right)~.
\end{eqnarray}\label{trans}
\end{subequations}
In Appendix~\ref{AppA}, we present a detail derivation for Eqs.~(\ref{trans}). 
It is found that the transformation in Eqs.~(\ref{trans}) is a conformal transformation, namely,
\begin{eqnarray}
  \textrm{d}\Psi^2+\sin^2\Psi\textrm{d}\Phi^2=\Omega(\Psi_0,\Phi_0)^2(\textrm{d}\Psi_0^2+\sin^2\Psi_0\textrm{d}\Phi_0^2)~, \label{confor}
\end{eqnarray}
where the conformal factor $\Omega$ can be given by 
\begin{eqnarray}
  \Omega(\Psi_0,\Phi_0)^2 = \frac{f_\text{o}-\omega^2 r_\text{o}^2\sin^2\theta_\text{o}}{\left(\sqrt{f_\text{o}}+\omega r_\text{o}\sin\theta_\text{o}\cos\Phi_0\sin\Psi_0\right)^2}~.
\end{eqnarray}
For the time-like 4-velocity of $u^{(\phi)}$, we have $\Omega^2 \propto f_\text{o}-\omega^2 r_\text{o}^2\sin^2\theta_\text{o} >0$, indicating that the transformation preserves orientation at any point on the celestial sphere.

For in/out-going observers with the 4-velocity $u^{(r)}=(\partial_t+v\partial_r)/\sqrt{f_\text{o}-v^2/f_\text{o}}$, we obtain the aberration formula as follows,
\begin{subequations}
  \begin{eqnarray}
  \Phi &=& \Phi_0 ~, \\
  \Psi &=& \arccos\left(\frac{f_\text{o}\cos\Psi_0-v}{f_\text{o}-v\cos\Psi_0} \right)~. 
\end{eqnarray} \label{trans2}
\end{subequations} 
Using Eqs.~(\ref{trans2}), we obtain the conformal factor in the form of
\begin{eqnarray}
  \Omega(\Psi_0,\Phi_0)^2 = \frac{f_\text{o}^2-v^2}{(f_\text{o}-v\cos\Psi_0)^2}~.
\end{eqnarray} 
In the case of $f_\text{o}=1$, the aberration formula in Eqs.~(\ref{trans2}) reduces to that in special relativity \cite{Einstein:1905ve,Terrell:1959zz,Penrose:1959vz}.

To formulate the relative speed between moving and static observers, one can introduce the covariant 3-speed as follows \cite{Gourgoulhon:2013gua},
\begin{eqnarray}
  \bar{v}\equiv \sqrt{\frac{\gamma^\ast u \cdot \gamma^\ast u}{(u_\text{stc}\cdot u)^2}}=\sqrt{1-\frac{1}{(u_\text{stc}\cdot u)^2}}~. \label{3speed}
\end{eqnarray}
Employing $u=u^{(\phi)}$ or $u=u^{(r)}$ in Eq.~(\ref{3speed}), we obtain 
\begin{eqnarray}
  \bar{v}^{(\phi)}=|\omega| r_\text{o} \sin\theta_\text{o}/\sqrt{f_\text{o}}~, &&
  \bar{v}^{(r)} = |v|/f_\text{o}~. \label{16}
\end{eqnarray}
Substituting Eqs.~(\ref{16}) into Eq.~(\ref{trans}) and Eq.~(\ref{trans2}), respectively, one might find that the function $f_\text{o}$ is eliminated. 
For aberration formula in Eqs.~(\ref{trans2}), it reduces to the form in special relativity \cite{Einstein:1905ve}.
In Appendix~\ref{AppD}, we verify that Eq.~(\ref{trans}) is also consistent with the result derived from local Lorentz transformation.

A key feature of the aberration formulas is their dependence solely on the observer's instantaneous velocity, independent of acceleration or trajectory, as shown in Eqs.~(\ref{trans}). For example, considering an observer moving along the $y$-axis in flat spacetime, the 4-velocity is given by
\begin{eqnarray}
  u=\frac{1}{\sqrt{1-\tilde{v}^2}}\left(1,\tilde{v}\sin\theta_\text{o}\sin\phi_\text{o},\frac{\tilde{v}\cos\theta_\text{o}\sin\phi_\text{o}}{r_\text{o}},\frac{\tilde{v}\cos\phi_\text{o}}{r_\text{o}\sin\theta_\text{o}}\right)~.
\end{eqnarray} 
The above 4-velocity reduces to $u^{(\phi)}$ with $\theta=\pi/2$, $\phi=0$ and $\tilde{v}=\bar{v}^{(\phi)}$. It indicates that, for an observer located at $(r_\text{o},\theta_\text{o},\phi_\text{o})=(r_\text{o},\pi/2,0)$, the aberration effect for an observer moving along the $y$-axis is indistinguishable from that predicted by that described by Eqs.~(\ref{trans}) with $f_\text{o}=1$.

\subsection{Apparent distortion from aberration effect \label{IIIB}}

\begin{figure} \centering
  \includegraphics[width=0.7\linewidth]{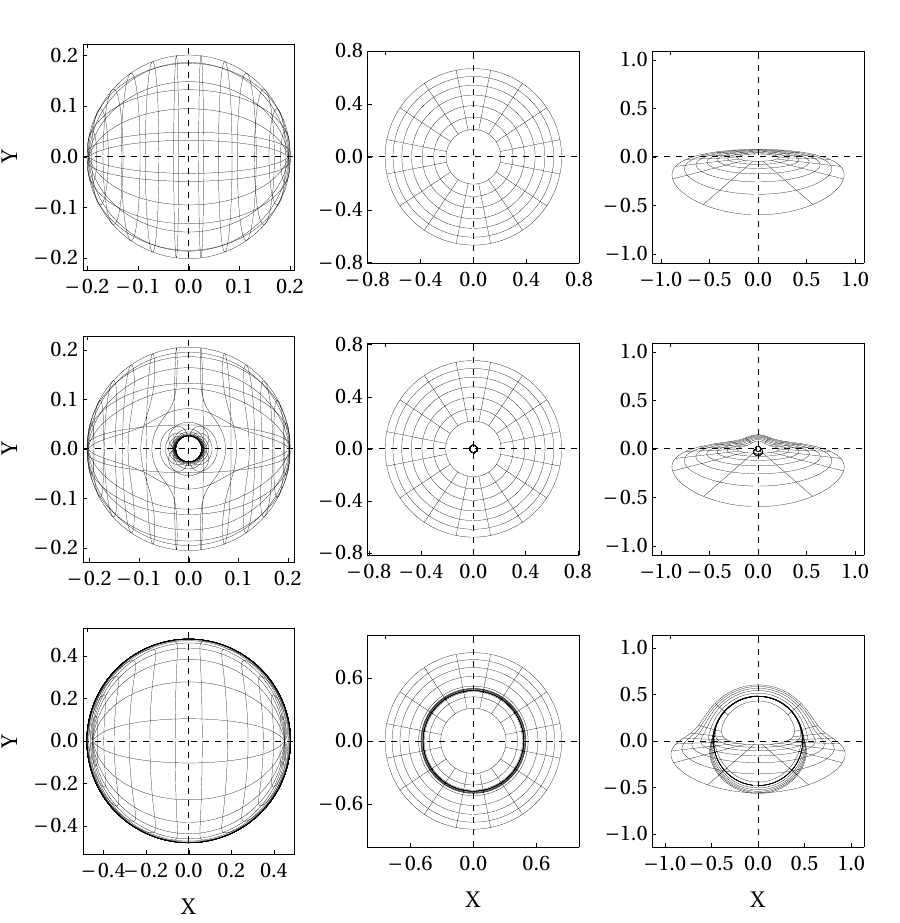}
  \caption{
    Optical appearance of spheres (left panels), face-on disks (middle panels), and disks with inclination angle $\theta_\text{o}=4\pi/9$ (right panels) with respect to static observer at distance $r_\text{o}=10l_g$. The radii of the spheres are set to be $2l_g$. The outer radii and inner radii of the disks are set to be $8.1 l_g$ and $2.1 l_g$, respectively. And we consider the flat spacetime, $l_g=20 r_g$ and $l_g=r_g$ for the top, middle and bottom panels, respectively.
  \label{F2}}
\end{figure}

\begin{figure*}  
  \centering
  \begin{subfigure}[b]{0.49\textwidth}
    \caption{$\bar{v}=0.5$ \label{F3a}}
  \includegraphics[width=1\linewidth]{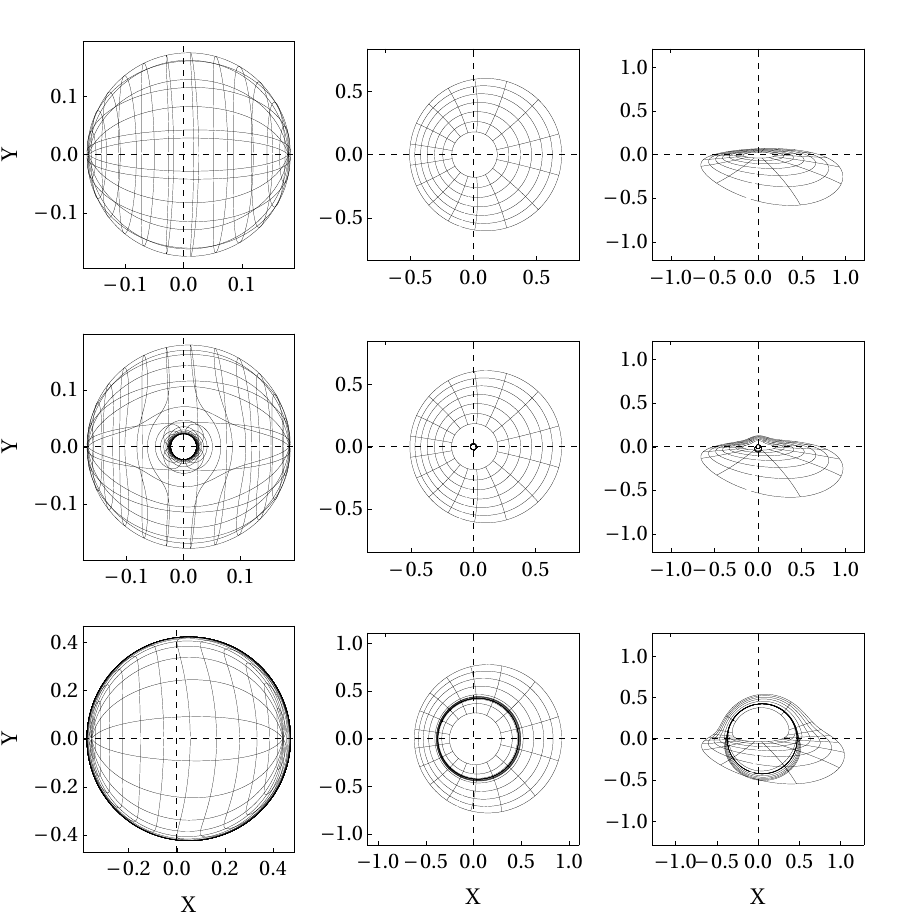}
    \end{subfigure} 
   \begin{subfigure}[b]{0.49\textwidth} 
    \caption{$\bar{v}=0.999$ \label{F3b}}
    \includegraphics[width=1\linewidth]{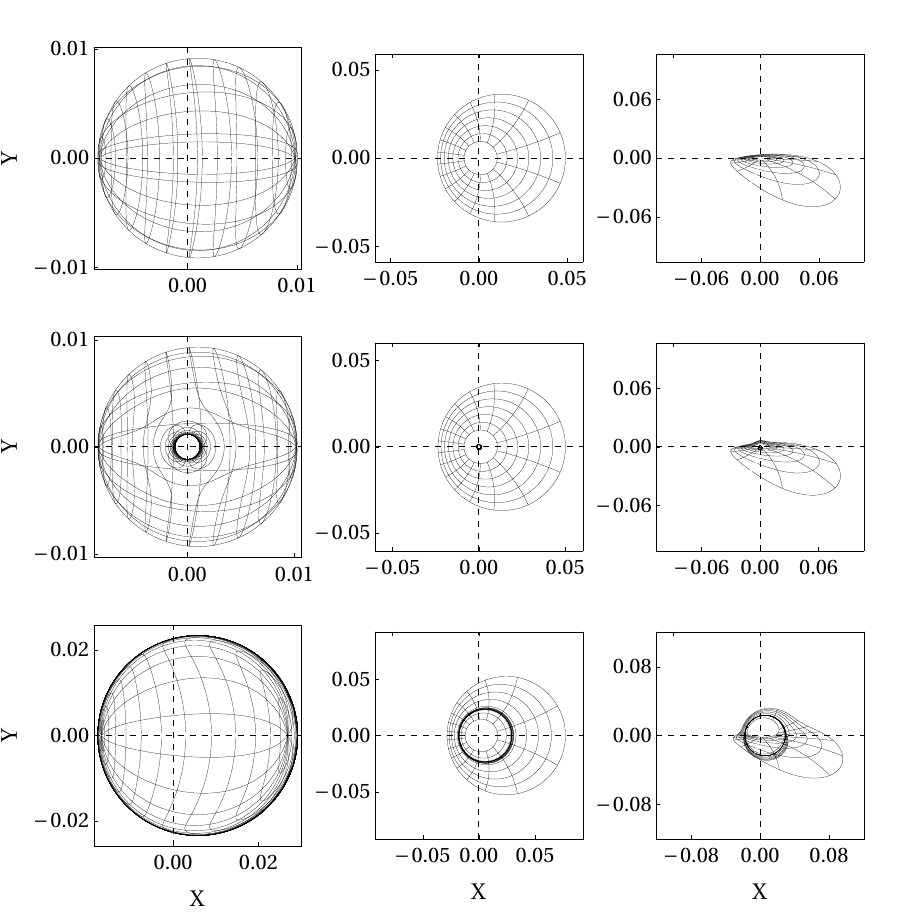}
      \end{subfigure}
  \caption{
    Optical appearance of spheres, face-on disks, and disks with inclination angle $\theta_\text{o}=4\pi/9$ with respect to moving observer at distance $r_\text{o}=10l_g$ and selected values of speed $\bar{v}^{(\phi)}$ ($\omega <0$). The radii of the spheres and disks are set to be the same as those in Fig.~\ref{F2}.\label{F3}} 
  
\end{figure*}

We numerically simulate the images of spheres and disks with respect to static observer in Fig.~\ref{F2}. The gravity can distort the images as shown in the middle and bottom panels of Fig.~\ref{F2}.  It is noticed that the images of the disks were not discussed in the seminal works \cite{Terrell:1959zz,Penrose:1959vz}. Here, we consider the disks in order to provide further insight into the properties of the aberration effect.
The images of spheres and disks for moving observers are presented in Fig.~\ref{F3}.   Through a comparison of Fig.~\ref{F2} and Fig.~\ref{F3}, we show that the apparent size of the sphere shrinks, and apparent center is shifted as the speed $\bar{v}$ increases. For the images of the spheres, it seems to exhibit an apparent rotation. However, it is found that the `rotation' angle has an upper limit as the speed $\bar{v}\rightarrow1$. For the images of the inclined disks, one might find the images get distorted. In the case of the face-on disks, the shape of circle within the disks remains unchanged. This occurs because the conformal transformation in Eq.~(\ref{confor}) preserves circles on the celestial sphere. 
 
All these results suggest that the Terrell-Penrose effect originating from the aberration effect is not appropriately described as a pure apparent rotation on the celestial sphere.

\section{Imaging a moving object\label{IV}}

In this section, we investigate the snapshots of a moving object seen by a static observer, to compare with the results in Sec.~\ref{III}. We employ slow-light ray tracing and time-series reconstruction.
Specifically, for a moving emission point with trajectory $\bm{x}_\text{s}(t_\text{s})$, the light ray emitted at $(t_\text{s},r_\text{s})$ and received by an observer at distance $r_\text{o}$ and time $t_\text{o}$ can be described by 
\begin{eqnarray}
  t_\text{o}-t_\text{s}&=& \pm_r \int_{r_\text{s}}^{r_\text{o}}\textrm{d}r\left\{\frac{r}{f\sqrt{r^2-\rho^2f}}\right\}\nonumber \\  &=:&\Delta T(r_\text{o},r_\text{s};\rho)~. \label{18}
\end{eqnarray}
For an extended body, the image is composed of the light emitted by all points on its surface. 
Specifically, an extended emission source can be modeled as a collection of discrete emission points, with the $i$-th point represented as $\bm x_{\text{s}}^{(i)}(t_{\text{s}}^{(i)})$. For an observer at fixed distance $r_\text{o}$ and the celestial coordinates measured at time $t_\text{o}$, the three-dimensional spatial geometry of the extended emission source is specified by the set of emission points that satisfy the following equation, namely,
\begin{eqnarray}
  t_\text{o} - t_\text{s}^{(i)}(\bm x_\text{s}^{(i)}) = \Delta T(r_\text{o},r_\text{s}^{(i)};\rho(\bm x_\text{o},\bm x_\text{s}^{(i)}))~. \label{19}
\end{eqnarray}
The  $\rho$ in Eq.~(\ref{19}) is determined by given $\bm x_\text{s}$ and $\bm x_\text{o}$, thus denoted $\rho(\bm x_\text{o},\bm x_\text{s}^{(i)})$. It is based on the one-to-one mapping illustrated in Eq.~(\ref{RT}). Given the trajectories $\bm x_{\text{s}}^{(i)}(t_{\text{s}}^{(i)})$ for the extended emission source, we can numerically solve Eq.~(\ref{19}) for the $\bm x^{(i)}_\text{s}$. The resulting solutions $\bm x^{(i)}_{\text{s},\ast}$ for all $i$ give the emission points seen by the observer at  $(t_\text{o},r_\text{o})$.

\subsection{Snapshots of moving spheres and disks in flat spacetime \label{IVA}}

\begin{figure} 
  \begin{subfigure}[t]{0.49\textwidth} \centering 
      \caption{\label{F4a}} 
  \includegraphics[width=0.8\linewidth]{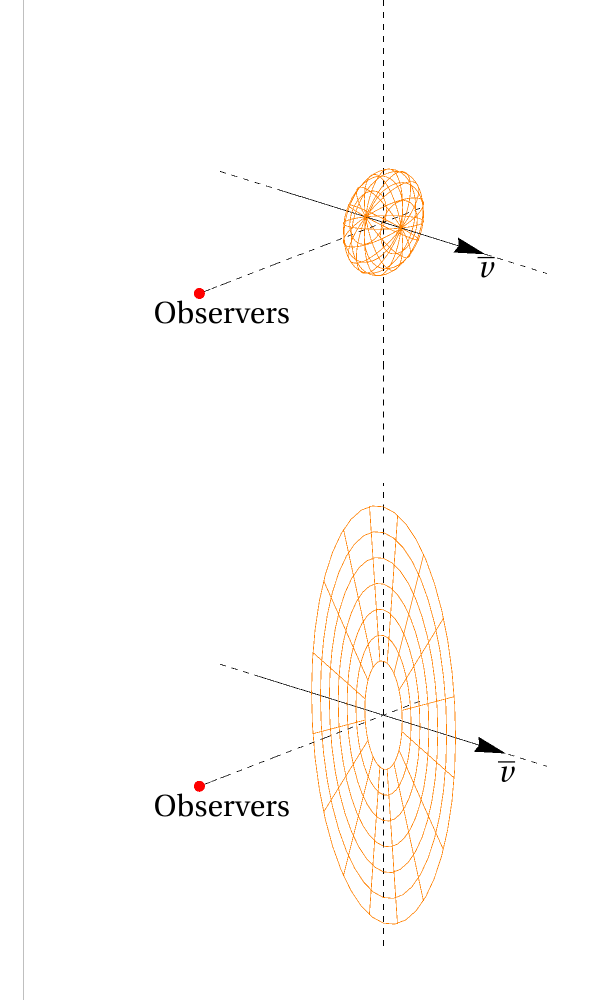}
  \end{subfigure}
  \begin{subfigure}[t]{0.49\textwidth} \centering 
    \caption{\label{F4b}} 
  \includegraphics[width=1\linewidth]{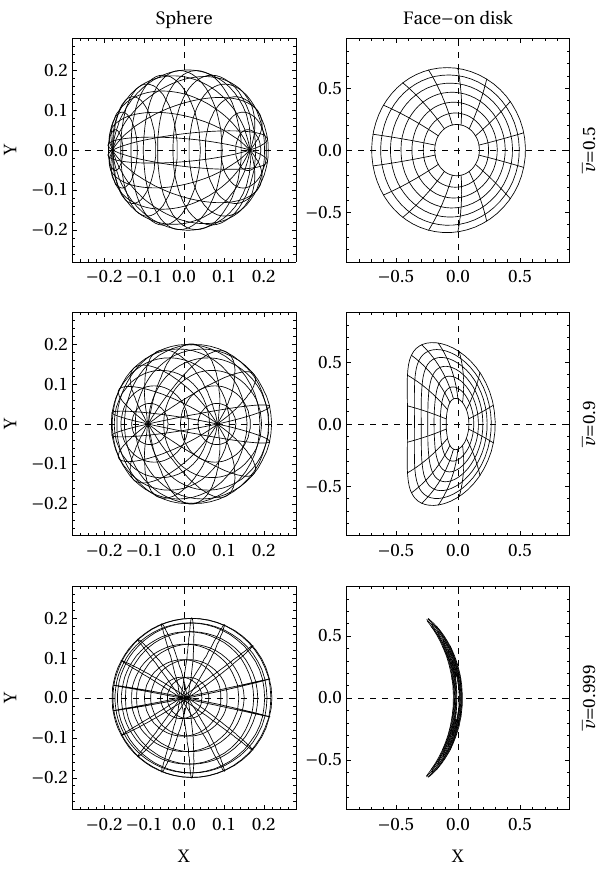}
  \end{subfigure}
  \caption{Panel (a): schematic diagram illustrating the Lorentz contraction of the moving sphere (top panel) and face-on disk (bottom panel). Panel (b): optical appearances of moving spheres (left panel) and moving face-on disks (right panel) in the speed of $\bar{v}$. The radii of the spheres and the face-on disks are set to be the same as those in the top panels of Fig.~\ref{F2}. All simulations are performed in flat spacetime. \label{F4}}
\end{figure}

Supposing that an extended body moves uniformly along the $y$-axis in the flat spacetime, the emission time of the $i$-th point is given by $t_\text{s}^{(i)} = \pm(y_\text{s}^{(i)}-y_{\text{s},0}^{(i)})/\bar{v}$, where $y_{\text{s},0}^{(i)}$ is the initial $y$-coordinate. Using Cartesian coordinates, we denote the position of the $i$-th emission point as $\bm x_\text{s}^{(i)} = (x^{(i)}_\text{s},y^{(i)}_\text{s},z^{(i)}_\text{s})$. Here, the $\bar{v}$ is the 3-speed defined in Eq.~(\ref{3speed}). For each index $i$, the coordinates $x_\text{s}^{(i)}$ and $z_\text{s}^{(i)}$ take constant values, since the motion is along the $y$-axis. The schematic diagrams for a  sphere and a disk moving along the $y$-axis are illustrated in Fig.~\ref{F4a}. Their 3-dimensional spatial shapes get distorted as a consequence of Lorentz contraction. It can be formulated as
\begin{subequations}
  \begin{eqnarray}
 \bm x_\text{s}^{(i)}(t_\text{s}^{(i)}) &=& (x^{(i)}_\text{s},y^{(i)}_\text{s,0}\pm\bar{v}t^{(i)}_\text{s},z^{(i)}_\text{s})~, \\
  0 &=&F(x^{(i)}_\text{s},y^{(i)}_\text{s}(t_\text{s}^{(i)}) \sqrt{1-\bar{v}^2} ,z^{(i)}_\text{s})~, \label{Feq}
\end{eqnarray} 
\end{subequations}
where Eq.~(\ref{Feq}) includes the effect of Lorentz contraction and $F(\bm x)=0$ defines the surface. For the surface of a sphere, we have $F(\bm x)= |\bm x-\bm x_0|^2 -R^2_\text{sphere}$, where $R_\text{sphere}$ is radius of the sphere and $\bm x_0$ is the initial coordinate. 

The optical appearances of moving spheres and disks are presented in Fig.~\ref{F4b}. From the images, the sphere appears to rotate without changing the shape. The motion of the sphere induces an apparent elongation along the $y$-axis due to the slow-light effect, which cancels the effect of Lorentz contraction. It is consistent with the suggestion by Terrell \cite{10.1119} and the illustrated in Ref.~\cite{Hornof:2024rzt}.  However, this result differs from that presented in the top-left panels of Fig.~\ref{F3a} and Fig.~\ref{F3b}. 
In the right panel of Fig.~\ref{F4b}, we further show that the apparent distortion of concentric circles in the disks varies with the speed $\bar{v}$. From the images, the face-on disk appears to get elliptical, differing from the results in the top-middle panels of Fig.~\ref{F3a} and Fig.~\ref{F3b}. The images of the disks in Fig.~\ref{F4b} explicitly show that the conformality on the celestial sphere is violated.

All these results indicate that the moving sources in the view of a static observer and static sources in the view of a moving observer do not produce identical images, even in flat spacetime.

\subsection{Apparent elongation influenced by gravity \label{IVB}}

Because the concept of rigidity for an extended body is ambiguous in curved spacetime, a direct comparison between the image of a moving sphere in flat spacetime and that in a black hole spacetime would be inappropriate. A flawed but illustrative example is provided in Appendix~\ref{AppE}. 

\begin{figure} 
\begin{subfigure}[t]{0.49\textwidth} \centering 
      \caption{\label{F5a}} 
  \includegraphics[width=0.5\linewidth]{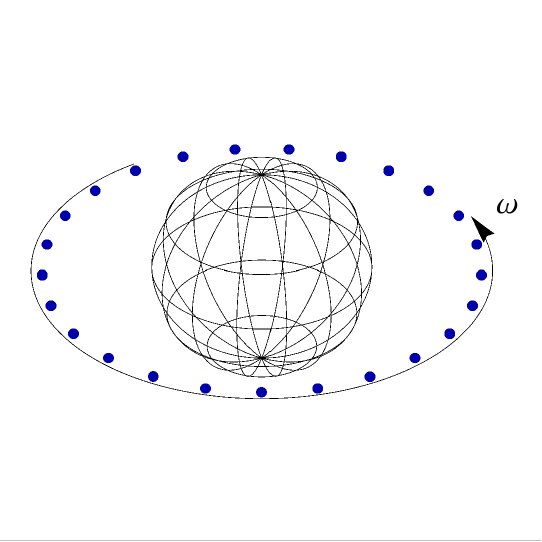}
  \end{subfigure}
  \begin{subfigure}[t]{0.5\textwidth} \centering 
    \caption{\label{F5b}} 
  \includegraphics[width=1\linewidth]{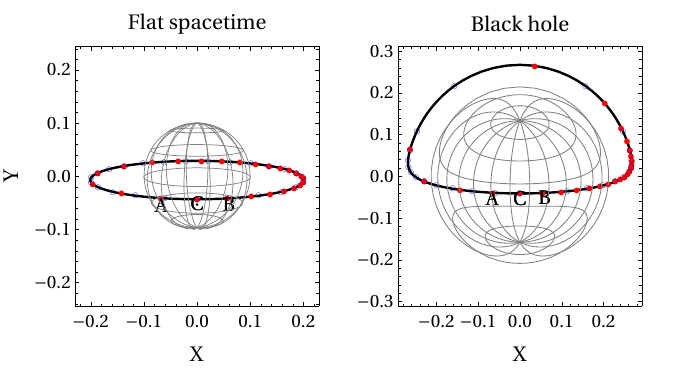}
  \end{subfigure}  
  \caption{Panel (a): emission sources are modeled as a collection of points corotating in equatorial circular orbits with orbital radius of $4l_g$. We present images of the points in flat spacetime [left panel of (b)] and in black hole spacetime with $l_g=r_g$ [right panel of (b)]. Red points and blue circles show the points at angular speed $\omega =0.15$ and $\omega =0$, respectively. Labels A, B, and C denote the points closest to the observer.  \label{F5}}
\end{figure}

\begin{figure} \centering
  \includegraphics[width=0.85\linewidth]{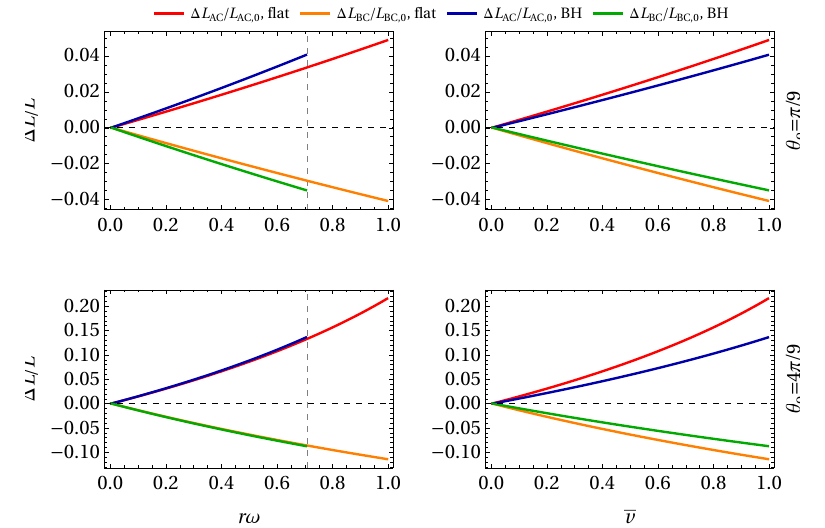}  
  \caption{Relative variation of angular distances $L_\text{AC}$ and $L_\text{BC}$ as functions of the coordinate speed $r\omega$ (left panels) and covariant 3-speed $\bar{v}$ (right panels) for selected inclination angles. The A, B, and C are the points shown in Fig.~\ref{F5}. 
   \label{F6}}
\end{figure}

Here, to avoid conceptual ambiguity, we consider a collection of moving points uniformly distributed along a circular orbit. The schematic diagram is shown in Fig.~\ref{F5a}. Physically, it corresponds to corotating hotspots in accretion disks.   Because the uniform orbital distribution is a consequence of dynamics of the accreting matter and the spherical symmetry of the spacetime \cite{Novikov:1973kta}, we can focus on the slow-light effect in snapshots of the hotspots and study  additional influence of gravity.  
Fig.~\ref{F5b} shows the images of the corotating hotspots in the absence and presence of gravity. It is found that the images of the points are non-uniformly distributed along the orbit, with a concentration in the region $Y\simeq0$ and $X>0$.
For quantitative analysis, we select three representative points, labeled A, B, and C in Fig.~\ref{F5b}. Angular distances between these moving points are denoted $L_\text{AC}$ and $L_\text{BC}$.
The slow-light effect is expected to produce variations in $L_\text{AC}$ and $L_\text{BC}$, formulated as $\Delta L_\text{BC}/L_\text{BC}\equiv L_\text{BC}/L_\text{BC,0}-1$ and $\Delta L_\text{AC}/L_\text{AC}\equiv L_\text{AC}/L_\text{AC,0}-1$, where $L_\text{BC,0}\equiv L_\text{BC}|_{\bar{v}=0}$ and $L_\text{AC,0}\equiv L_\text{AC}|_{\bar{v}=0}$. Fig.~\ref{F6} shows the relative variation of the angular distance $\Delta L/L$ as a function of the orbital speed.
It is found that $L_\text{AC}$ is contracted, while $L_\text{BC}$ is elongated. 
This arises because points A and B are imaged at emission times earlier than that of point C. 
It also causes the variation $|\Delta L/L|$ to increase with the angular speed $\omega$. And for large $\theta_\text{o}$, the apparent distance between points A and B is elongated due to $|\Delta L_\text{AC}/L_\text{AC}|>|\Delta L_\text{BC}/L_\text{BC}|$ as shown in the bottom panels of Fig.~\ref{F6}.
In the presence of gravity, the absolute speed could be ambiguous. It can be quantified with either the coordinate speed, $r\omega$, or the covariant 3-speed, $\bar{v}$. In both cases, it shows that gravity has an additional influence on the $\Delta L/L$. 
Specifically, supposing that the covariant 3-speed $\bar{v}$ is physically relevant (the right panels of Fig.~\ref{F6}), we find that gravity suppresses the apparent elongation.

\section{A static sphere seen by a moving observer versus a moving sphere seen by static observer \label{V}}

\begin{figure} \centering 
  \includegraphics[width=0.8\linewidth]{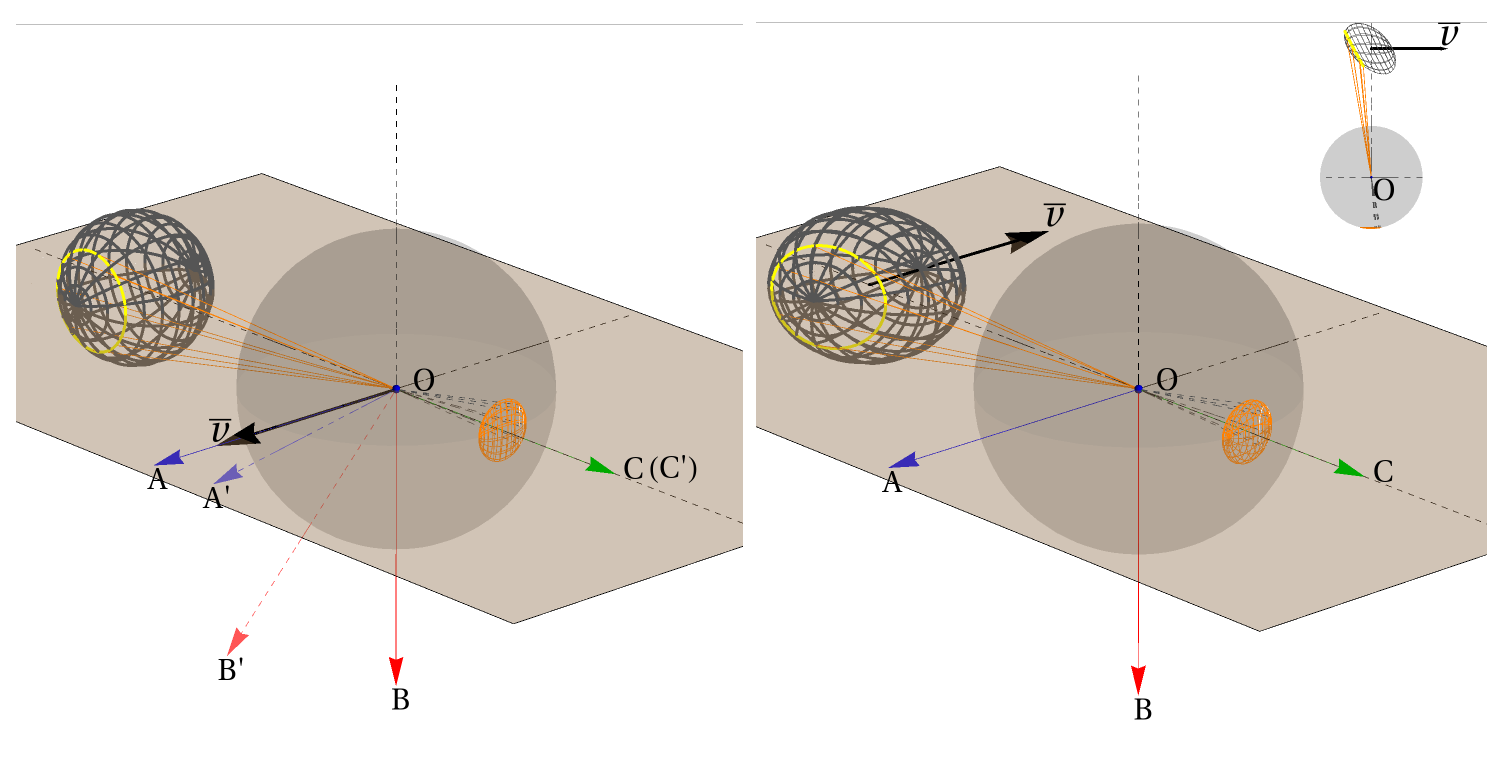}  
  \caption{Left panel: imaging a static sphere on the observer's celestial sphere for observer in motion (system $\rm X$-$\rm A'B'C'$) or not (system $\rm O$-$\rm ABC$). The observer is located at O. The set of vectors \{$\overrightarrow{\text{OA}'}$,$\overrightarrow{\text{OB}'}$,$\overrightarrow{\text{OC}'}$\} are the spatial parts of $\{k^{(\phi')}, k^{(\theta')}, k^{(r')} \}$. Right panel: imaging a moving sphere on the observer's celestial sphere with respect to a static observer.  The sphere gets distorted in the space, due to the Lorentz contraction and slow-light effect. In both cases, we set $\bar{v}=0.5$, the radius of the sphere is $2l_g$, and the observer's distance to the sphere's center is $10l_g$. The ray-tracing simulations are performed in flat spacetime. ($l_g= 10^4 r_g \gg r_g$). \label{F7}}
\end{figure}

In this section, we compare the apparent distortion of static spheres originating from the aberration effect in Sec.~\ref{IIIB} [case i)] and the snapshot of a moving sphere seen by a static observer in Sec.~\ref{IVA} [case ii)]. Notably, the most tricky feature of case i) is that the apparent distortion of the images depends solely on the observer's spatial position and instantaneous velocity, as formulated by Eqs.~(\ref{11}), with no dependence on the observer's acceleration and trajectory. Specifically, for an observer in uniform motion along the $y$-direction with 4-velocity $u^{(y)}=(\partial_t+(vy/r)\partial_r + (v x_0/r^2)\partial_\phi)/\sqrt{1- v^2}$ in flat spacetime ($f_\text{o}=1$), we derive the aberration formula using Eqs.~(\ref{A2}), namely,
\begin{subequations}
  \begin{eqnarray}
  \Phi &=& \arctan\left( \frac{(r_\text{o}- v y)\sin\Phi\sin\Psi}{v x_0(1-\cos\Psi)+(r_\text{o}-v y)\cos\Phi\sin\psi} \right) ~, \\
  \Psi &=& \arccos\left(1-\frac{\left(1-\cos \Psi_0\right) \left(1- v^2\right)}{ (r_\text{o}-v y)(r_\text{o}-v y\cos\Psi + v x_0 \cos\Phi\sin\Psi}\right)~, 
\end{eqnarray} \label{transy}
\end{subequations} 
where observer moves along the $y$-direction with speed $v$, and $r_\text{o}=\sqrt{x_0^2+y^2}$. At the observer's location of $y=0$, the aberration formula in Eq.~(\ref{transy}) reduces to Eq.~(\ref{trans}) with $\theta=\pi/2$ and $\omega=r_\text{o} v$. It indicates that the results presented in Sec.~\ref{IIIB} are equivalent to the case where the observer moves along the $y$-direction at the location $(r_\text{o},\theta_\text{o},\phi_\text{o})=(r_\text{o},\pi/2,0)$. Hence, the top panels of Fig.~\ref{F3} and the right panel of Fig.~\ref{F4} represent the contrasting cases of a sphere at rest with an observer moving along the negative $y$-direction [case i)], and a sphere moving along the positive $y$-direction for a static observer [case ii)], respectively.  Although the relative velocity between the sphere and observer is identical in these cases, our results show that the resulting images are not identical.
\begin{table*} \centering
  \caption{Summary of the distinct features displayed in the top panels of Fig.~\ref{F3} and the right panel of Fig.~\ref{F4}. \label{T0}}
  \begin{ruledtabular}
  \begin{tabular}{l|ccc} 
    & Sphere & Face-on disk &  \\  \hline
    Case i): moving observer & \begin{tabular}{l}
  1. No rotation \\
  2. Circular silhouette  \\ 
  3. Distorted interior
\end{tabular} & \begin{tabular}{l}
  1. Right-shifted center \\
  2. Circle preserving 
\end{tabular} \\ \hline
    Case ii): moving object & Apparent rotation  & Get elliptical&  \\
  \end{tabular}
\end{ruledtabular}
\end{table*}
We list these distinct features in Tab.~\ref{T0}.
To show the robustness of our results, the ray tracing and imaging procedures are presented in Fig.~\ref{F7}. 

As a result of aberration effect, the celestial sphere is redefined using a set of new reference light rays, $\{k^{(\phi')}, k^{(\theta')}, k^{(r')}\}$ given in Eq.~(\ref{lightk}). (The celestial coordinate system is transformed from $\rm O$-$\rm ABC$ to $\rm O$-$\rm A'B'C'$, as shown in the left panel of Fig.~\ref{F7}). Because the static sphere emits light persistently, the paths of light rays from the sphere to the observer are independent of the observer's velocity. Since these paths of light rays remain unchanged, all the image distortion from the aberration effect is formulated by a local transformation on the celestial sphere at O. 

On the other hand, due to the Lorentz contraction and the slow-light effect, the moving sphere at the observed time $t_\text{o}$ is deformed in the 3-dimensional space as shown in the left panel of Fig.~\ref{F7}. (The distorted sphere in the 3-dimensional space is formulated by a collection of points $\bm x_{\text{s},\ast}^{(i)}$, which are the solutions of Eq.~(\ref{19}).) It directly results in the corresponding image of the sphere appearing rotated.

The discrepancy in these cases can also be illustrated by examining the specific aspect in the following.
In case i), if the sphere is opaque, emission points on the back side of the sphere would never be seen by the moving observer. In case ii), it allows the observer to see the emission points on the back side of the moving sphere. Thus, one can not expect that the spheres in these two cases produce the identical images.

\section{Conclusions and discussions  \label{VI}}

We examined the Terrell-Penrose effect for two contrasting cases: i) of static emission sources in the view of moving observers, and ii) moving emission sources in the view of static observers. The images for the emission sources in these cases are distinct, even in flat spacetime.  Specifically,  in case i), adopting an operationalist approach, we revisited that the apparent distortion of the images can be formulated as a conformal transformation. It is consistent with the results in the seminal works by Terrell and Penrose \cite{Terrell:1959zz,Penrose:1959vz}. In case ii), the image of the sphere appears rotated, which is consistent with the picture of the Terrell rotation \cite{10.1119,Hornof:2024rzt}. We summarize the image difference for sphere and face-one disk in Tab.~\ref{T0}. Here, the most important distinction we found is that in case i), the image distortion can be described by a conformal transformation on the celestial sphere, whereas in case ii), it does not have such a result. 

Given that the sources and observers are located in different gravitational environments, the images in cases i) and ii) should be inherently distinct, thereby encoding the spacetime geometry in principle. In flat spacetime, the image distortion is governed by the aberration formula in case i), whereas in case ii), it is determined by Lorentz contraction and slow-light effect. 
However, in curved spacetime, Lorentz contraction is ill-defined, and a rigorous comparison between these two cases can not be established. In this context, we restrict our study to the aberration and slow-light effects as affected by gravity.
In case i), the extension of the aberration formula to curved spacetime is trivial: it reduces to the same expression as that in special relativity by replacing the speed with the covariant 3-speed, $\bar{v}$ [Eq.~(\ref{3speed})].  
If one accepts, as shown in the seminal works \cite{Terrell:1959zz,Penrose:1959vz}, that the Terrell-Penrose effect is fully described by aberration formulas, then our study successfully extends this effect to curved spacetime, as detailed in Sec.~\ref{IIIA}.
If one believes that the Terrell-Penrose effect must include the setup of a moving object, as in case ii), then we note that we only show that gravity has an additional influence on the slow-light effect in Sec.~\ref{IVB}. In this sense, a complete description of the Terrell-Penrose effect is not provided.

Imaging an object on the celestial sphere differs from describing it in a conventional inertial frame.
Specifically, the reception time and position on the celestial sphere are defined by when and where light signals are received by the observer. In contrast, a standard inertial frame is defined by a network of synchronized clocks at rest throughout space, which establish a global time and record the position of the clock. The key distinction is that these two frames employ different operational definitions of time: one is asynchronous, based on light propagation, while the other is synchronous across space. This explains why the Lorentz contraction of a sphere is not directly visible to observers, because the image of a moving object is collectively determined by its historical positions \cite{10.1119,Hornof:2024rzt}. In this sense, the results for imaging a moving object [case ii)] seemingly align with our physical intuition.

Based on the principles of relativity, pioneers considered case ii) equivalent to treating the sphere as being at rest and the observer as moving [case i)], and thus, the aberration formulas were applied \cite{Penrose:1959vz,Terrell:1959zz}. Our study also adopts this framework to simulate the results, as shown in Sec~\ref{III}. However, the aberration formula is demonstrated to be independent of the observer's historical position. Specifically, according to Eq.~(\ref{11}), the position determined by celestial coordinates depends only on the instantaneous relative velocity of the observer. This resulting discrepancy on the images between case i) and ii) has not been previously anticipated or elucidated by the pioneers \cite{Penrose:1959vz,Terrell:1959zz}. Perhaps, a plausible explanation for this discrepancy is that the aberration formula can not fully account for the image distortion for a moving observer, as another physical effect is systematically missing. 

In the context of the Terrell-Penrose effect, the observable is the astrometric position defined on the observer's celestial sphere, as given in Eqs.~(\ref{CS}). This is a purely geometric, achromatic effect. However, this is idealized, in physical reality, light emitted from the object's surface must experience a frequency shift. In particular, the observed frequency is sensitive to both the gravitational field and the object's motion, indicating that it can encode additional physical information. It would be of great interest to study the frequency shift in our two contrasting cases, and to examine their equivalence. However, since frequency shifts are beyond the scope of the  Terrell-Penrose effect, we did not consider them in the present study. The frequency-dependent feature becomes more non-trivial when light rays are replaced by timelike particles in black hole spacetime. In this case, the ray-tracing simulation for imaging sources around a black hole can be related to a particle-scattering problem in  black hole spacetime. We believe that it would also be worthwhile to extend the study of aberration and slow-light effects to particle scattering.
\

{\it Acknowledgments.} This work is supported by the National Natural Science Foundation of China under grants No.~12305073 and No.~12347101.
The author thanks Prof. Hongbao Zhang and Dr. Jie Zhu for useful discussions. 

\bibliography{ref} 
 

\appendix
\section{Intermediate results of Eqs.~(\ref{trans}) and Eqs.~(\ref{trans2}) \label{AppA}}

By making use of Eq.~(\ref{6}), the reference light rays in Eqs.~(\ref{lightk}) can be evaluated to be
 \begin{subequations}
   \begin{eqnarray}
    k^{(r')} & = & k^{(r)} ~,\\
    k^{(\phi')} & = & k^{(\phi)} - \frac{u \cdot k^{(\phi)}}{u \cdot k^{(r)}} \langle
    k^{(\phi)}, k^{(r)} \rangle k^{(r)} ~,\\
    k^{(\theta')} & = & k^{(\theta)} + \frac{(u \cdot k^{(\theta)}) (u \cdot k^{(r)})
    (\langle k^{(\theta)}, k^{(\phi)} \rangle - \langle k^{(r)}, k^{(\phi)}
    \rangle \langle k^{(\theta)} , k^{(r)} \rangle)}{k^{(r)} \cdot
    k^{(\phi)} (1 + \langle k^{(r)} , k^{(\phi)} \rangle)} k^{(\phi)} \nonumber \\
    && - \Bigg( \frac{(u \cdot k^{(\theta)}) (u \cdot k^{(\phi)}) \langle k^{(r)},
    k^{(\phi)} \rangle (\langle k^{(\theta)}, k^{(\phi)} \rangle - \langle
    k^{(r)}, k^{(\phi)} \rangle \langle k^{(\theta)}, k^{(r)}
    \rangle)}{k^{(r)} \cdot k^{(\phi)} (1 + \langle k^{(r)} , k^{(\phi)}
    \rangle)}  + \frac{u \cdot k^{(\theta)}}{u \cdot k^{(r)}} \langle
    k^{(\theta)}, k^{(r)} \rangle \Bigg) k^{(r)} ~. \nonumber \\
  \end{eqnarray} \label{A1}
 \end{subequations}
With these reference light rays, the celestial coordinates $(\Phi,\Psi)$ for moving observers can be derived from 
  \begin{subequations}
    \begin{eqnarray}
    \cos \Phi \sin \Psi & \equiv &  \langle k, k^{(\phi')} \rangle \nonumber \\ &=& \frac{\langle k, k^{(\phi)} \rangle - \langle k, k^{(r)} \rangle
    \langle k^{(\phi)}, k^{(r)} \rangle}{\sqrt{1 - \langle k^{(\phi)}, k^{(r)}    \rangle^2}}~,\\
    \sin \Phi \sin \Psi& \equiv &    \langle k, k^{(\theta')} \rangle \nonumber \\ &=& \mathcal{N}^{-1}\Bigg(1 + (\langle k^{(r)}, k^{(\phi)} \rangle + 1) \left(
    \frac{\langle k, k^{(\theta)} \rangle - \langle k^{(\theta)}, k^{(r)}
    \rangle \langle k, k^{(r)} \rangle + \langle k^{(\theta)}, k^{(r)} \rangle -
    1}{1 - \langle k^{(\theta)}, k^{(r)} \rangle - \langle k^{(\theta)},
    k^{(\phi)} \rangle + \langle k^{(r)}, k^{(\phi)} \rangle} \right)  \nonumber \\
    &&  +  \left( \frac{\langle k, k^{(\phi)}
    \rangle - \langle k^{(r)}, k^{(\phi)} \rangle \langle k, k^{(r)} \rangle +
    \langle k^{(r)}, k^{(\phi)} \rangle - 1}{1 - \langle k^{(\theta)}, k^{(r)}
    \rangle - \langle k^{(\theta)}, k^{(\phi)} \rangle + \langle k^{(r)},
    k^{(\phi)} \rangle} \right)  \left( \frac{\langle k^{(\theta)}, k^{(\phi)} \rangle - 
    \langle k^{(r)},
    k^{(\phi)} \rangle \langle k^{(\theta)}, k^{(r)} \rangle}{\langle
    k^{(r)}, k^{(\phi)} \rangle - 1} \right)\Bigg) ~,\nonumber \\ \\
    \cos \Psi & \equiv &  \langle k, k^{(r)} \rangle~,
  \end{eqnarray} \label{A2}
  \end{subequations}
where
\begin{eqnarray}
  \mathcal{N}^{2}=1-\frac{2}{1-\langle k^{(r)}, k^{(\phi)} \rangle}\left(1 - \frac{(1-\langle k^{(\theta)}, k^{(\phi)} \rangle)(1-\langle k^{(\theta)}, k^{(r)} \rangle)(1+\langle k^{(r)}, k^{(\phi)} \rangle)^2}{(1-\langle k^{(\theta)}, k^{(r)} \rangle+\langle k^{(r)}, k^{(\phi)} \rangle-\langle k^{(\theta)}, k^{(\phi)} \rangle)^2}\right)~.
\end{eqnarray}
  Using the inner products of the null vectors in Tab.~\ref{T1}, we obtain the explicit expressions of Eqs.~(\ref{trans}) and Eqs.~(\ref{trans2}).
\begin{table*} \centering
  \caption{Results of redshifts and product between 4-velocities of light rays with respect to observers $u^{(*)}$ introduced in Eq.~(7)\label{T1}}
  \begin{ruledtabular}
  \begin{tabular}{c|ccc} 
      & $u^{(\phi)}$  &  $u^{(r)}$ & \\ 
    \hline
    $1+z$ & $\frac{f_\text{o}  + f_\text{o} \omega r_\text{o}\sin\theta_\text{o} \cos \Phi_0 \sin \Psi_0}{\sqrt{f_\text{o}^2  
    - f_\text{o} \omega^2 r_\text{o}^2\sin^2\theta_\text{o}}}$ & $\frac{f_\text{o}-v\cos\Psi_0}{\sqrt{f_\text{o}^2-v^2}}$ & \\
    $1+z^{(\phi)}$ & $\frac{f_\text{o} + \sqrt{f_\text{o}} \omega r_\text{o}\sin\theta_\text{o}}{\sqrt{f_\text{o}^2  - f_\text{o} \omega^2 r_\text{o}^2\sin^2\theta_\text{o}}} $ &$\frac{f_\text{o}-v}{\sqrt{f_\text{o}^2-v^2}}$  & \\
    $1 + z^{(\theta)}$ & $\frac{f_\text{o}}{\sqrt{f_\text{o}^2   - f_\text{o} \omega^2 r^2    \sin^2 \theta_\text{o}}}$ & $\frac{f_\text{o}}{\sqrt{f_\text{o}^2-v^2}}$ & \\
    $1+z^{(r)}$ & $\frac{f_\text{o}}{\sqrt{f_\text{o}^2   - f_\text{o} \omega^2 r^2    \sin^2 \theta_\text{o}}}$ & $\frac{f_\text{o}}{\sqrt{f_\text{o}^2-v^2}}$& \\
     $\langle k^{(\phi)}, k^{(r)} \rangle$ & $\frac{\omega r_\text{o} \sin \theta_\text{o}}{\sqrt{f_\text{o}}}$ & $-\frac{v}{f_\text{o}}$ &  \\
     $\langle k^{(\phi)}, k^{(\theta)} \rangle$ & $\frac{\omega r_\text{o} \sin \theta_\text{o}}{\sqrt{f_\text{o}}} $ & $\frac{v^2}{f_\text{o}^2}$ & \\
     $\langle k^{(\theta)}, k^{(r)} \rangle$ & $\frac{r^2
    \omega^2 \sin^2 \theta_\text{o}}{ f_\text{o} }$  & $-\frac{v}{f_\text{o}}$ &\\
     $\langle k, k^{(\phi)} \rangle$ & $\frac{\sqrt{f_\text{o}} \sin \Psi_0  \cos \Phi_0+r \omega  \sin \theta_\text{o} }{\sqrt{f_\text{o}}+r \omega    \sin \theta_\text{o}  \sin \Psi_0 \cos \Phi_0 }$  & $1-\frac{(f_\text{o}^2-v^2)(1-\cos\Phi_0\sin\Psi_0)}{f_\text{o}(f_\text{o}-v\cos\Psi_0)}$ & \\
     $\langle k, k^{(\theta)} \rangle$ &  $1 -  \frac{(f_\text{o}  -  \omega^2    r_\text{o}^2\sin^2\theta_\text{o})(1 - \sin \Phi_0 \sin \Psi_0)}{f_\text{o} + \sqrt{f_\text{o}} \omega  r_\text{o}\sin\theta_\text{o} \cos \Phi_0 \sin \Psi_0}
   $  & $1-\frac{(f_\text{o}^2-v^2)(1-\cos\Phi_0\sin\Psi_0)}{f_\text{o}(f_\text{o}-v\cos\Psi_0)}$   &\\
     $\langle k, k^{(r)} \rangle$ & $1 -   \frac{(f_\text{o}  -  \omega^2
    r_\text{o}^2\sin^2\theta_\text{o})(1 - \cos \Psi_0)}{f_\text{o}  + \sqrt{f_\text{o}} \omega  r_\text{o}\sin\theta_\text{o} \cos \Phi_0 \sin \Psi_0} $  & $\frac{f_\text{o}\cos\Psi_0-v}{f_\text{o}-v\cos\Psi_0}$ & \\ 
  \end{tabular}
  \end{ruledtabular}
\end{table*}

\section{Aberration formula in special relativity for moving observers along $y$-axis \label{AppB}}

Associating the 4-velocity $u^{(y)}=(\partial_t+\lambda\partial_\theta)/\sqrt{1-\lambda^2 r^2}$ and $f=1$ with the Eqs.~(\ref{A2}), we obtain aberration formula as follows,
\begin{subequations}
  \begin{eqnarray}
  \Phi &=& \arctan\left(\frac{\sqrt{f_\text{o}} \left( \sqrt{f_\text{o}} \sin \Phi_0\sin \Psi_0+\lambda  r_\text{o} \left(\cos \Phi    _0\sin \Psi_0+ \cos \Psi_0-1\right)\right)}{f_\text{o} \cos \Phi_0 \sin \Psi_0-\sqrt{f_\text{o}}\lambda   r_\text{o} \sin \Phi   _0 \sin \Psi_0-\lambda ^2 r_\text{o}^2 \left(\cos   \Psi_0-1\right)}\right) ~, \\
  \Psi &=& \arccos\left(1-\frac{\left(1-\cos \Psi_0\right) \left(f_\text{o}-\lambda ^2 r_\text{o}^2\right)}{f_\text{o}-\lambda  \sqrt{f_\text{o}} r_\text{o} \sin   \Phi_0 \sin \Psi_0}\right)~. 
\end{eqnarray} \label{trans3}
\end{subequations} 
Based on Eqs.~(\ref{trans3}), the conformal factor takes the form of
\begin{eqnarray}
  \Omega(\Phi_0,\Psi_0)^2=\frac{f-\lambda^2r_\text{o}^2}{\left(\sqrt{f_\text{o}}-\lambda r_\text{o} \sin\Phi_0\sin\Psi_0\right)^2}~.
\end{eqnarray}
The calculation follows the same procedure as described in Appendix~\ref{AppA}.

\section{The infinitesimal transformations and space symmetries}
The celestial sphere can be mathematically treated as the Riemann sphere to study the space symmetries \cite{Penrose:1959vz}. Namely, using stereographic projection, 
\begin{eqnarray}
  z = e^{i\Phi_0}\cot\left(\frac{\Psi_0}{2}\right)~, \label{25}
\end{eqnarray}
the Riemann sphere are mapped onto the complex plane of $z$. The global conformal group, consisting of the transformations acting on the extended complex plane, constitutes the symmetry group of the Riemann sphere. The generators span sub-algebras of the Witt algebra, namely
\begin{eqnarray}
  l_n = - z^{n+1}\partial_z~, &\hspace{0.5cm} &
  \bar{l}_n = - \bar{z}^{n+1} \partial_{\bar{z}}~, \label{C22}
\end{eqnarray}
where $n=-1$, $0$, $1$. The Lie algebra structure is defined by the commutation relations, $[ l_n,l_m]= (n-m)l_{n+m}$ and $[l_n,\bar{l}_m]=0$.

We have shown that the aberration formulas in Eqs.~(\ref{trans}), Eqs.~(\ref{trans2}) and Eqs.~(\ref{trans3}) are conformal transformations. 
It would be interesting to explore the relationship between the local symmetries and the types of moving observers.
The infinitesimal transformations on the celestial sphere can be written in the form of $\Psi=\Psi_0+\varepsilon \xi^\Psi$ and $\Phi=\Phi_0+\varepsilon \xi^\Phi$, where the $\varepsilon$ is parameter of the transformation and the infinitesimal vector $\xi$ can be given by
\begin{eqnarray}
  \xi & \equiv & \left( \frac{\partial \Phi}{\partial \varepsilon}
  \frac{\partial}{\partial \Phi} + \frac{\partial \Psi}{\partial
  \varepsilon} \frac{\partial}{\partial \Psi} \right)_{\varepsilon = 0}
  ~. \label{27}
\end{eqnarray}
Associating Eqs.~(\ref{trans}), (\ref{trans2}) and (\ref{trans3}) with Eq.~(\ref{27}), we obtain
\begin{subequations}
  \begin{eqnarray}
  \xi^{(r)} & = & \frac{ \sin \Psi_0}{f} \frac{\partial}{\partial \Psi_0}~,\\
  \xi^{(\phi)} & = & - \frac{ r \sin \theta}{\sqrt{f}} \Bigg( 2 \cos \Phi_0
  \sin^2 \left( \frac{\Psi_0}{2} \right) \frac{\partial}{\partial \Psi_0} \nonumber \\ && +
  \sin \Phi_0 \tan \left( \frac{\Psi_0}{2} \right) \frac{\partial}{\partial
  \Phi_0} \Bigg)~,\\
  \xi^{(\theta)} & = & \frac{ r}{\sqrt{f}} \Bigg( 2 \sin \Phi_0 \sin^2 \left(
  \frac{\Psi_0}{2} \right) \frac{\partial}{\partial \Psi_0} \nonumber \\ && + \left( 1 - \cos
  \Phi_0 \tan \left( \frac{\Psi_0}{2} \right) \right) \frac{\partial}{\partial
  \Phi_0} \Bigg)~.
\end{eqnarray} \label{28}
\end{subequations}
Based on stereographic projection in Eq.~(\ref{25}) and generators in Eqs.~(\ref{C22}), the infinitesimal vectors in Eqs.~(\ref{28}) can be rewritten as
\begin{subequations}
  \begin{eqnarray}
  \xi^{(r)}  &=&  \frac{1}{f} (l_0 + \bar{l}_0)~,\\
  \xi^{(\phi)}  &=&  - \frac{ r \sin \theta}{\sqrt{f}} (l_{- 1} + \bar{l}_{-
  1})~, \\
  \xi^{(\theta)}  &=&  - \frac{i    r}{\sqrt{f}} (l_0 - \bar{l}_0 - l_{-
  1} + \bar{l}_{- 1})~. 
\end{eqnarray}\label{29}
\end{subequations}
It indicates that on the complex plane, the $\xi^{(r)}$ generates dilations, the $ \xi^{(\phi)}$ generates translations, and the $\xi^{(\theta)}$ is a linear combination of the rotation and translation generators. 

\begin{figure} \centering
  \centering
  \includegraphics[width=0.7\linewidth]{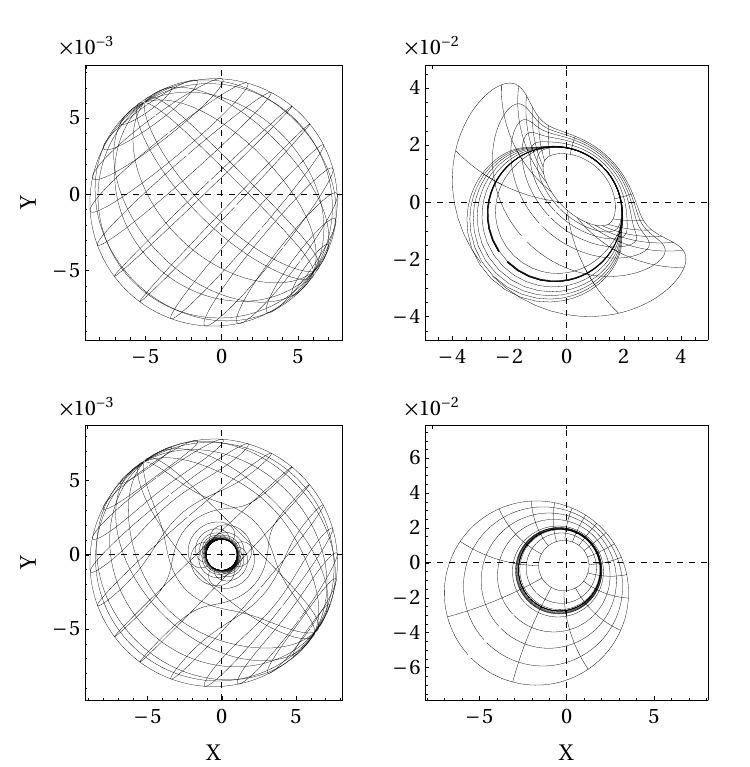}
  \caption{Optical appearance of spheres and inclined and face-on accretion disk with respect to $\theta$-motion observers at distance of $10l_g$ and speed of $\bar{v}=0.999$ and $\lambda<0$. The radii of the spheres and disks are set to be the same as those in Fig.~\ref{F2}. We consider flat spacetime for the top-left panel, $l_g=20r_g$ for  the bottom-left panel, and $l_g=r_g$ for the right panels.  \label{F8}}
\end{figure}

We have $\xi^{(\phi)}\propto f^{-1/2}$, $\xi^{(\theta)} \propto f^{-1/2}$ and $\xi^{(r)} \propto f^{-1}$, indicating distinct scaling between radial and angular directions in black hole spacetime ($f\neq 1$).
Moreover, $\xi^{(\theta)}$ differs from $\xi^{(\phi)}$ by an additional rotation generator. This conclusion is also evident in the apparent distortions of the spheres and disks in Fig.~\ref{F8}. It arises from the orthogonalization scheme in Eqs.~(\ref{lightk}), which preserves the $r$-$\phi$ planes. %
It is mathematically unavoidable and should be understood as one of the properties of relativity.

\section{Aberration formulas and local Lorentz transformation \label{AppD}}

Using local Lorentz transformation in a curved spacetime, one can obtain the aberration formula as follows \cite{Terrell:1959zz,Einstein:1905ve,Gourgoulhon:2013gua},
\begin{subequations}
  \begin{eqnarray}
  \Psi&=&\Psi_0~,\\
  \cos\Psi&=& \frac{\cos\Psi_0-V}{1-V\cos\Psi_0}~, 
\end{eqnarray}\label{C1}
\end{subequations}
where the $V$ is the local 3-velocity defined in the local frame \cite{Gourgoulhon:2013gua}. The celestial coordinates $(\Phi_0,\Psi_0)$ and $(\Phi,\Psi)$ have equivalent status, as the inverse transformation of Eqs.~(\ref{C1}) can be obtained by interchanging $\Phi\leftrightarrow \Phi_0$, $\Psi\leftrightarrow \Psi_0$, and $V\leftrightarrow -V$.  In this sense, a rotation of the celestial coordinates about the $y$-axis by $\pi/2$ is given by
\begin{subequations}
  \begin{eqnarray}
  \left( \begin{array}{c}
   \cos \Psi'\sin \Psi' \\ 
   \sin \Psi' \sin \Psi' \\
   \cos \Psi'
 \end{array} \right) &=& \left( \begin{array}{ccc}
   &  & \hspace{0.1cm}1 \\ 
    & 1 &  \\
  -1 &  & 
 \end{array} \right) \left( \begin{array}{c}
   \cos \Psi\sin \Psi \\ 
   \sin \Psi \sin \Psi \\
   \cos \Psi
 \end{array} \right)~, \nonumber \\ \label{C2a} \\
   \left( \begin{array}{c}
   \cos \Psi'_0\sin \Psi'_0 \\ 
   \sin \Psi'_0 \sin \Psi'_0 \\
   \cos \Psi'_0
 \end{array} \right) &=& \left( \begin{array}{ccc}
    &  & \hspace{0.1cm}1 \\ 
    & 1 &  \\
  -1 &  & 
 \end{array} \right) \left( \begin{array}{c}
   \cos \Psi_0\sin \Psi_0 \\ 
   \sin \Psi_0 \sin \Psi_0 \\
   \cos \Psi_0
 \end{array} \right)~. \nonumber \\ 
\end{eqnarray} \label{C2}
\end{subequations}
Based on Eqs.~(\ref{C2}), the aberration formulas can be evaluated to be
\begin{subequations}
  \begin{eqnarray}
    \tan\Phi' &=& \frac{\sin\Phi_0'\sin\Psi_0'\sqrt{1-V^2}}{\cos\Phi_0'\sin\Psi_0'-V}~, \\
    \cos\Psi' &=& \frac{\cos\Psi_0'\sqrt{1-V^2}}{1-V\cos\Phi_0'\sin\Psi_0'}~.
  \end{eqnarray}\label{C3}
\end{subequations}
By Substituting $V=-\omega r_\text{o}\sin\theta_\text{o}/\sqrt{f_\text{o}}$, one might find that the expressions of Eqs.~(\ref{C3}) are different from that of Eqs.~(\ref{trans}). 
The fixed point of the transformation in Eqs.~(\ref{C3}) is $(\Phi_0,\Psi_0)=(0,\pi/2)$, while the fixed point of the transformation in Eqs.~(\ref{trans}) is $(0,0)$.

From the shifted fixed point, we realized that the celestial coordinates $(\Phi,\Psi)$ and $(\Phi_0, \Psi_0)$ should be rotated about the $y$-axis by different rotation angles, namely,

\begin{subequations}
  \begin{eqnarray}
  \left( \begin{array}{c}
   \cos \Psi'\sin \Psi' \\ 
   \sin \Psi' \sin \Psi' \\
   \cos \Psi'
 \end{array} \right) &=& \left( \begin{array}{ccc}
   \cos\Theta & 0 & \sin\Theta \\ 
   0 & 1 & 0 \\
  -\sin\Theta & 0 & \cos\Theta
 \end{array} \right) \left( \begin{array}{c}
   \cos \Psi\sin \Psi \\ 
   \sin \Psi \sin \Psi \\
   \cos \Psi
 \end{array} \right)~,  \\
   \left( \begin{array}{c}
   \cos \Psi'_0\sin \Psi'_0 \\ 
   \sin \Psi'_0 \sin \Psi'_0 \\
   \cos \Psi'_0
 \end{array} \right) &=& \left( \begin{array}{ccc}
   \cos\Theta_0 & 0 & \sin\Theta_0 \\ 
   0 & 1 & 0 \\
  -\sin\Theta_0 & 0 & \cos\Theta_0
 \end{array} \right) \left( \begin{array}{c}
   \cos \Psi_0\sin \Psi_0 \\ 
   \sin \Psi_0 \sin \Psi_0 \\
   \cos \Psi_0
 \end{array} \right)~, 
\end{eqnarray} \label{B1}
\end{subequations}
where $\Theta = \arccos \langle k^{(\phi)}, k^{(r)} \rangle$ and $\Theta_0 =\pi/2$. Specifically, we have $\langle k^{(\phi)}, k^{(r)} \rangle = -V$ for the radial observer $u^{(r)}$.  The angle $\Theta\neq \pi/2$ originates from the aberration effect. By making use of Eqs.~(\ref{B1}) and Eqs.~(\ref{C1}), we obtain
\begin{subequations}
  \begin{eqnarray}
    \tan \Phi' & = & \frac{\sin \Phi \sin \Psi}{- V
  \cos \Phi \sin \Psi + \cos \Psi \sqrt{1 -
  V^2}} \nonumber \\ &=& \frac{  \sin \Phi_0 \sin \Psi_0}{
  \cos \Psi_0 - V (\cos \Phi_0 \sin \Psi_0 + 1)}  \nonumber\\  
  & = & \frac{ \sin \Phi_0' \sin \Psi_0'}{ \cos
  \Phi_0' \sin \Psi'_0 - V (1 - \cos \Psi'_0)} ~,\\
    \cos \Psi' & = & - \cos \Phi \sin \Psi \sqrt{1 -
  V^2} - V \cos
  \Psi \nonumber \\ &=& 1 - \frac{(1 + \cos \Phi_0 \sin \Psi_0) \left( 1 -
  V^2 \right)}{1 - V
  \cos \Psi_0} \nonumber \\
  & = & 1 - \frac{(1 - \cos \Psi'_0) \left( 1 -
  V^2 \right)}{1 - V
  \cos \Phi'_0 \sin \Psi'_0}~.
  \end{eqnarray} \label{B2}
\end{subequations}
By setting $V = -\omega r_\text{o} \sin\theta_\text{o}/\sqrt{f_\text{o}}$, Eqs.~(\ref{B2}) reduces to the form of Eqs.~(\ref{trans}). We thus demonstrate that the aberration formula obtained using reference light rays is consistent with that based on local Lorentz transformation. Here, different fixed points can be selected under different conventions for the coordinate system, and thus are not important.

\section{Another example for illustrating apparent distortion influenced by gravity \label{AppE}}

\begin{figure} \centering   
\includegraphics[width=0.8\linewidth]{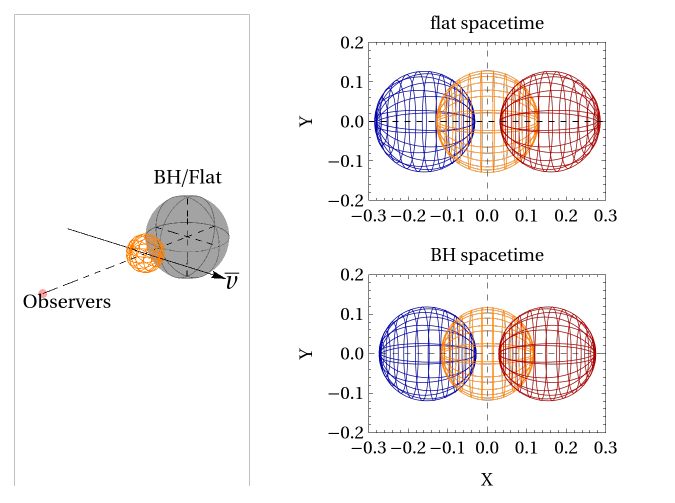}  
\caption{Left panel: schematic diagram of a sphere moving along the $y$-axis with constant  speed $\bar{v}$ in front of a black hole. Right panels: optical appearances of the static sphere located along $y$-axis with the observer's distance to the coordinate center of $10 l_g$. The radius of the spheres is $0.9 l_g$ and the distance from the orange sphere's center to the coordinate center is $3l_g$. In the top-right and bottom-right panels, we set $l_g=10^4 r_g$ and $l_g=r_g$, respectively. Here, we have neglected the Lorentz contraction. \label{F9}}
\end{figure}

\begin{figure} \centering 
\includegraphics[width=0.7\linewidth]{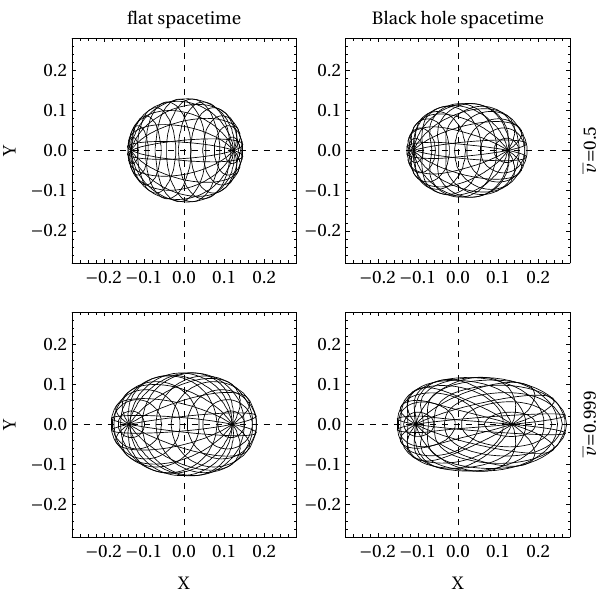}  
\caption{The optical appearances of a moving sphere without Lorentz contractions in flat spacetime (left panels) and in black hole spacetime with $l_g=r_g$ (right panels). The configurations of the spheres and the locations of the observers are shown in Fig.~\ref{F9}. 
\label{F10} 
}
\end{figure}

Suppose that a sphere moves along the $y$-axis in front of a black hole, as illustrated in the left panel of Fig.~\ref{F9}. The Cartesian coordinate is heuristically defined as 
\begin{eqnarray}
  (x,y,z) = (r\sin\theta\cos\phi,r\sin\theta\cos\phi,r\cos\theta)~.
\end{eqnarray}
Each point on the moving sphere has the same covariant 3-speed $\bar{v}$ along the $y$-axis, giving the trajectory of the $i$-th point in the form of
\begin{eqnarray}
  t_\text{s}^{(i)}= \frac{1}{\bar{v}}\int^{y_s^{(i)}}_{y_{s,0}^{(i)}} \mathrm{d}y\left\{\sqrt{\frac{1}{f}\left(\frac{1}{f}-1\right)\left(\frac{y}{r}\right)^2+\frac{1}{f}}\right\}~,
\end{eqnarray}
where $r=\sqrt{(x_s^{(i)})^2+y^2+(z_s^{(i)})^2}$.
In the right panel of Fig~\ref{F9}, we present images of the static sphere located along the $y$-axis. 
This shows that gravity has only a slight influence on the images of static spheres.  By performing a ray-tracing simulations, we present snapshots of the moving sphere in Fig.~\ref{F10}. Comparing the left and right panels of Fig.~\ref{F10}, it shows that the gravity significantly influences on the snapshots of the spheres in motions, in contrast to the images of static spheres. 

\end{document}